\documentclass[preprint]{aastex}

\shorttitle{STIS Observations of HR 4796A}
\shortauthors{Schneider et al.}

\begin{document}
\renewcommand{\topfraction}{1.0}
\renewcommand{\bottomfraction}{1.0}
\renewcommand{\textfraction}{0.3}
\textwidth 6.25in
\textheight 9in

\title{STIS Imaging of the HR 4796A Circumstellar Debris Ring}

\author{G. Schneider}
\affil{Steward Observatory, The University of Arizona, 933 N. Cherry Avenue, Tucson, AZ 85721}
\email{gschneider@as.arizona.edu}

\author{A. J. Weinberger}
\affil{Dept. of Terrestrial Magnetism,
Carnegie Institution of Washington, 5241 Broad Branch Road NW, Washington, DC 20015}
\email{weinberger@dtm.ciw.edu}

\author{E. E. Becklin}
\affil{Department of Physics and Astronomy, University of California
at Los Angeles, Los Angeles, CA 90095}
\email{becklin@astro.ucla.edu}

\author{J. H. Debes}
\affil{Dept. of Terrestrial Magnetism,
Carnegie Institution of Washington, 5241 Broad Branch Road NW, Washington, DC 20015}
\email{debes@dtm.ciw.edu}

\author{B. A. Smith}
\affil{Institute for Astronomy, University of Hawaii, 2680 Woodlawn Drive, Honolulu, HI 96822}
\email{basmith@ifa.hawaii.edu}

\begin{abstract}

We have obtained high spatial resolution imaging observations of the HR
4796A circumstellar debris dust ring using the broad optical response of
the Hubble Space Telescope Imaging Spectrograph in coronagraphic mode.
We use our visual wavelength observations to improve upon the earlier
measured geometrical parameters of the ring-like disk.  Two significant
flux density asymmetries are noted: (1) preferential forward scattering
by the disk grains and (2) an azimuthal surface brightness anisotropy
about the morphological minor axis of the disk with corresponding
differential ansal brightness. We find the debris ring offset from the
location of the star by $\sim$1.4 AU, a shift insufficient to explain the
differing brightnesses of the NE and SW ansae simply by the 1/$r^2$
dimmunition of starlight.  The STIS data also
better quantify the radial confinement of the starlight-scattering
circumstellar debris, to a characteristic region $<$~14 AU in
photometric half-width, with a significantly steeper inner truncation
than outward falloff in radial surface brightness.  The inferred spatial
distribution of the disk grains is consistent with the possibility of
one or more unseen co-orbital planetary-mass perturbers, and the colors
of the disk grains are consistent with a collisionally evolved
population of debris, possibly including ices reddened by radiation
exposure to the central star.
\end{abstract}


\section{Introduction}

HR 4796A (spectral type A0V) is the primary member of a binary system
with a parallactic distance of 72.8 $\pm$ 1.7 pc in the newly revised
Hipparcos catalog by \citet{vanLeeuwen2007}. It is interesting to note
that the distance has changed by more than 1$\sigma$ from the
determination of 67.1 $^{+3.6} _{-3.2}$ pc in the original Hipparcos
catalog of \citet{Perryman1997}.  Its age, based upon that of its
7\farcs7 distant M2.5V coeval companion, is estimated at 8 $\pm$ 2 Myr
\citep{Stauffer1995}.

The likely existence of a dusty debris disk around HR 4796A was inferred by
\citet{Jura1991} from its thermal infrared signature. He and collaborators
\citep{Jura1995} later suggested grain growth into larger particles at
distances 40--200 AU from the central star.  These predictions were borne out
from 10--20~$\mu$m mid-IR observations by \citet{Koerner1998} and
\citet{Jayawardhana1998} indicative of dust emission in a radial region
bounded at 55--80 AU (now 60 -- 87 with the revised distance).  Coronagraphic
observations by \citet{Schneider1999}, with the Hubble Space Telescope's ({\it
  HST}) Near Infrared Camera and Multi-Object Spectrometer (NICMOS) at 1.1 and
1.6 $\mu$m (spatial resolutions 110 and 160 mas, respectively), directly
imaged the dust, revealing a ring, 1\farcs05 (76 AU) in radius, with a width,
as measured by its photometric FWHM, of $\lesssim$17 AU (now $\lesssim$18.5
AU). The NICMOS imagery also suggested that the NE side of the disk was
brighter than the SW; a similar brightness asymmetry was seen in mid-IR
observations by \citet{Telesco2000} and modeled by
\citet{Wyatt1999}. Preferential ``forward'' scattering by the dust grains was
considered in models of the disk by \citet{Augereau1999}.

We obtained high spatial ($\approx$ 70 mas) resolution optical images of the
HR 4796A debris ring using the Space Telescope Imaging Spectrograph (STIS) in
its coronagraphic imaging mode.  From these data we examine in more detail the
spatial distribution of the ring particles as inferred from spatially resolved
imaging photometry along (and around) the ring.  We confirm, and better
quantify, the geometrical parameters defining the ring system and also measure
two anisotropies in the photometric properties along the ring.  Finally, we
take this opportunity to discuss some of the details, intricacies, and
methodologies employed in the reduction, processing, and analysis of HST/STIS
coronagraphic imaging data.  These will be the highest angular resolution
visible-light images of the ring for some time.  HST's Advanced Camera for
Surveys coronagraph provides a 0\farcs9 radius image plane mask, but 
has an effective inner working angle of $>$1\farcs2, limited mainly by point
spread function (PSF) subtraction residuals \citep[e.g.][]{Clampin2003}. The HR 4796A
disk is too small to be seen with that system.

\section{Observations} \label{Observations}

A program of high spatial resolution imaging of the HR 4796A
circumstellar debris ring was carried out using STIS in its
coronagraphic imaging mode as part of HST general observer program 8624
(PI: Weinberger).  With a foreknowledge of the ring geometry, an
optimized imaging sequence was planned and executed using broadband
optical coronagraphy (unfiltered spectral passband covering $\approx$ 2000 --
10500~\AA\ with a pivot wavelength ($\lambda_p$) of 5752~\AA, where
$\lambda_p$ is defined such that F$_\lambda$=F$_\nu
c$/$\lambda^2_p$, and FWHM=4330~\AA \footnote{Pivot wavelength and FWHM values
 taken from Space Telescope Science Data Analysis System Synphot software} ). Observations of HR 4796A were obtained at two epochs,
2001 Feb 17 and 2001 Feb 27, at spacecraft orientations differing by
16$\degr$, thereby rotating the debris ring orientation at the detector
focal plane with respect to the STIS coronagraphic occulting wedge.  At
each epoch the orientation of the telescope was selected to
simultaneously place the HR 4796A ring ansae at a maximum angular
distances from the STIS coronagraphic ``A'' wedge and from the stellar
diffraction spikes (see Table \ref{tab_logobs}).

\begin{deluxetable}{ccccccc}
\tablecolumns{4}
\tablewidth{0pc}
\tablecaption{Observing Log for HR 4796A and HR 4748\tablenotemark{a}\label{tab_logobs}}
\tablehead{
\colhead{Visit} 
        &\colhead{UT Date} 
                 &\colhead{Star}
                                 &\colhead{Orient\tablenotemark{b}}
                                         &\colhead{UT Start}
                                                    &\colhead{Exp. time per image (s)}
                                                          &\colhead{\# of images}}
\startdata
11   &17 Feb 2001  &HR 4796A    &282.058  &17:38:16 &6    &18\\
A1   &17 Feb 2001  &HR 4748    &283.272   &18:04:51 &4.2  &18\\
12   &27 Feb 2001  &HR 4796A     &298.057 &05:53:18 &6    &18\\
A2   &27 Feb 2001  &HR 4748     &281.420  &07:16:01 &4.2  &18\\
\enddata
\tablenotetext{a}{Additional exposure specific details may be obtained
  from the HST archive at http://archive.stsci.edu}
\tablenotetext{b}{Position angle of image Y axis (East of North).}
\end{deluxetable}

The STIS 50CCD aperture has an image scale of 50.70 mas~pixel$^{-1}$ and,
unfiltered, provides a spatial resolution of $\approx$ 70 mas for
stellar sources of B-V color index 0.0 (such as HR 4796A and HR 4748);
thus slightly undersampling the PSF.  When combined (in later processing)
re-registered images at two field orientations provide better sampling
of the PSF and also facilitate discriminating optical artifacts in the
HST/STIS PSF from true circumstellar features.

Observations of a brighter star of nearly identical spectral type, HR
4748 (to serve as a point spread function (PSF) subtraction template)
were obtained at each epoch.  To minimize temporal variations in the
HST/STIS PSF structure due to ``breathing'' of the telescope assembly
\citep{Schneider2001} and small instabilities in the projected position
of the STIS wedge onto the CCD focal plane, at each epoch both HR 4796A
and HR 4748 were observed during the same uninterrupted target
visibility period.  This entailed a small spacecraft slew of
1.3\degr\ to HR 4748 following our HR 4796A observations (maintaining
essentially the same spacecraft Sun angle for both pointings) with
subsequent guide star and coronagraphic target acquisitions occurring
without an intervening Earth occultation.

On-board autonomous target acquisitions placed the stars on the mid-line
of the the standard coronagraphic ``A'' wedge where its tapered width
narrows to 1\farcs0.  Each target was then moved 7\farcs4 along the
wedge mid-line to a location where the wedge is only 0\farcs63 wide.
Target placement at this ``non-standard'' location along the wedge was
used to maximize the visible portion of the small HR 4796A ring. The
target and PSF template star placements behind the occulting wedge were
all within the nominal (27 mas, 3 $\sigma$) combined STIS acquisition
and pointing precision of the telescope.

The exposure sequences are given in Table \ref{tab_logobs} and
summarized here.  With the star occulted by the coronagraphic wedge,
images were obtained with the unfiltered STIS CCD for total integration
times of 108s for HR 4796A and 75.6s for HR 4748 at each epoch. Each
individual image was 6 s (HR 4796A) or 4.2 s (HR 4748) in duration, and
the 18 individual images were used for cosmic ray rejection and
compensation. The individual exposures were planned to yield 72\% of
full-well depth at edge of the wedge based upon ground-based photometry
of the target stars and estimates of the anticipated instrumentally
scattered and diffracted stellar light.

Except for the field orientations and integration times, the HR 4796A
and HR 4748 exposures taken at the two epochs were executed in an
identical manner, as detailed in Table 1.  To reduce lost time in
operational overheads, we employed a sub-array readout mode of 80 rows
centered on the target position. As a result, the known M-star companion
to HR 4796A \citep{Jura1993} does not appear in the images.

In addition to the coronagraphic images, target acquisition images of
both stars (at both epochs) were obtained in the F25ND3
filter. Broadband images of these bright stars could not be obtained
without saturating the detector.

\section{Data Reduction, Processing, and Post-Processing}
\label{Data}

\subsection{Calibration} \label{Calibration}

\begin{figure}
\plotone{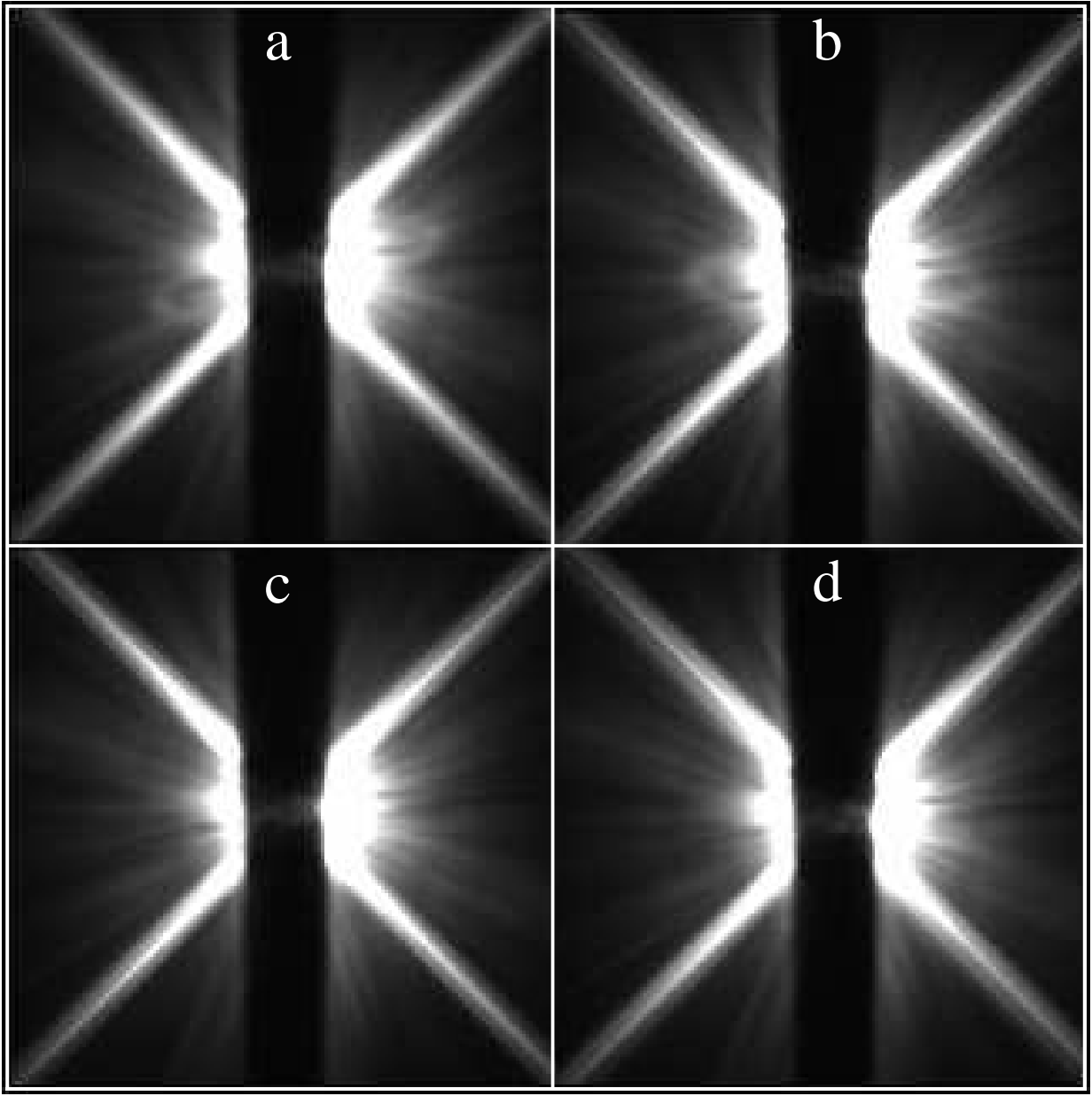}
\caption{Panels a and b show HR 4796A in calibrated STIS 50CCD coronagraphic
  images from the two visits and panels c and d show the PSF reference star HR
  4748 in the corresponding visits. All panels are 3\farcs9 $\times$ 3\farcs9
  and centered on the location of the star.  The position angles of the $+$y
  axis are 282.06$^\circ$ east of north in (a) and 298.06$^\circ$ in (b). The
  HR 4748 images have been scaled in brightness by a factor of 0.7 to match
  those of the paired contemporaneous observations of HR
  4796A. \label{fig_rawimages}}
\end{figure}

The raw CCD frames were processed with the STSDAS {\it calstis} software
task\footnote{available from: http://www.stsci.edu/hst/stis/software/}
using library flat field and dark current reference files from STScI's
calibration data base and nearly contemporaneous hot/bad pixel maps
provided by the STIS Instrument Definition Team (IDT).  At each epoch
the individual images were combined for cosmic-ray rejection using the
STSDAS {\it crreject} task and pixel values were converted to
instrumental count rates (counts/second). Unrejected bad/hot pixels
remaining in the combined count rate images were replaced by
two-dimensional Gaussian weighted interpolation (with a weighting radius
of 3 pixels) of neighboring good pixels.  A residual frame-dependent
fixed bias offset in the images was removed by subtracting a constant to
force the background to zero far from location of the star.  The
calibrated count rate images of HR 4796A and HR 4748 (re-normalized to
the flux density of HR 4796A) at both epochs (and hence different field
orientations) are shown in Figure \ref{fig_rawimages}.  Though embedded
in a wash of scattered and diffracted light from the core of the stellar
PSF not rejected by the coronagraph, the HR 4796A ring is clearly
visible in both HR 4796A frames (panels a and b).  The HR 4796A field
(and hence the orientation of the ring) is rotated 16\degr\ between the
two epochs, whereas the underlying structure of the HST+STIS PSF remains
fixed on the detector.  The rotational invariance of the stellar
component of the PSF is apparent when comparing this to the
contemporaneous observations of the reference star HR 4748 panels c and
d of Figure \ref{fig_rawimages}.

\subsection{PSF Subtraction} \label{Subtractions}

For each epoch, the calibrated HR 4748 images were simultaneously
flux-scaled, astrometrically registered, and subtracted from their
corresponding HR 4796A images with sub-pixel re-registration and
resampling accomplished by sync-function apodized cubic convolution
interpolation \citep{Park1983} as implemented in the NICMOS IDT's IDP3
software \citep{Stobie06}. To make the best subtractions, we measured
and iteratively minimized  the PSF-subtracted diffraction
spikes at radii of 15--30 pixels from the star (a region unaffected by
disk flux but close to the radii at which the disk is brightest), as a
function of the position and flux scaling of
the PSF template star. This was done separately for each visit-pair of
target-template subtractions. Other means of optimizing the PSF
subtractions, included flux scaling using the total energy along the
diffraction spikes beyond a radius of 30 pixels while demanding that the
median flux density of the background region to be statistically zero,
give very similar results.

The best HR 4748:HR 4796A 50CCD band flux-density scaling was found to
be 0.705. This is in statistical agreement with the F25ND3 ratio found
from the target acquisition images. We estimate the uncertainty in this
scaling as $\pm$ 0.002 by trying different minimization radii and
widths.  For the 17 February visit, the determined positional offsets
of the PSF star to HR 4796 were $\Delta x$=-0.360 and $\Delta y$=+0.388
pixels.  For the 27 February visit, the offsets were $\Delta x$=-0.075
and $\Delta y$=0.000. We estimate the uncertainty in all target:PSF star
position offsets as $\pm$ 0.020 pixels in both orthogonal directions
from examination of contours of $\chi^2$ as a function of different
positions.  The relatively large ($\approx$ 18 mas) offsets on 17 Feb
resulted from a target miscentering in the autonomous acquisition
process that, unfortunately, introduced additional optical artifacts
near the edge of the occulting wedge.  The positions, scale-lengths, and
amplitudes of the optical artifacts due to mis-centering under the wedge
are different from those which are induced by ``breathing'' and are in
discussed in \S \ref{Artifacts}.

\begin{figure}
\plotone{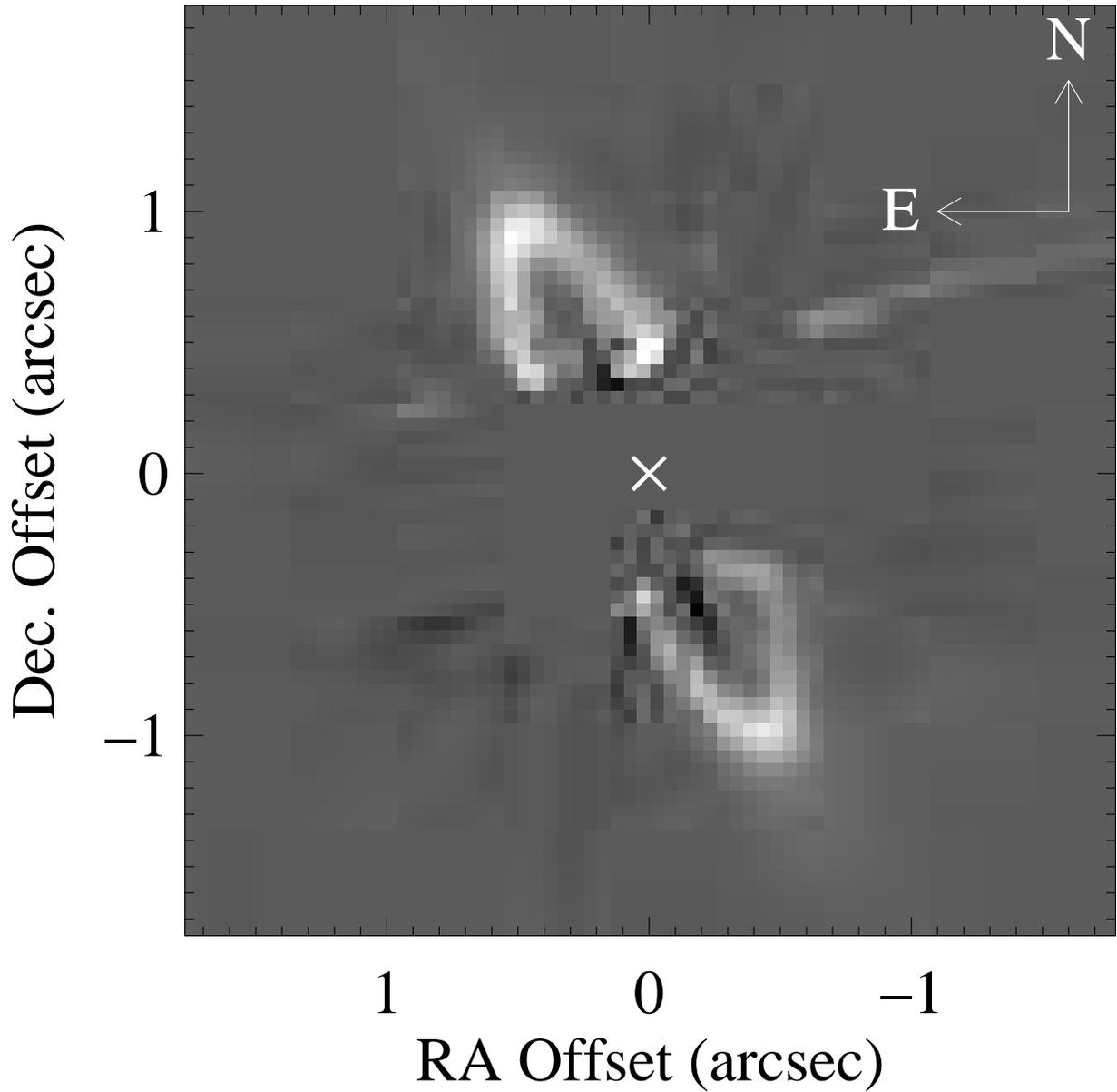}
\caption{Final STIS 50CCD PSF-subtracted coronagraphic image of the HR
  4796A circumstellar debris ring. The $\times$ marks the location of the
  star, as determined from the diffraction spikes in the unsubtracted
  image. \label{fig_finalimg}}
\end{figure}

Our final image of the HR 4796A circumstellar ring is the average of the two
visits; it is shown in Figure \ref{fig_finalimg} and used for all further
analysis in this paper.  When averaging the two PSF-subtracted images, we masked
out pixels unobserved due to the occulting wedge, pixels affected by the
diffraction spikes and pixels degraded by wedge-edge artifacts in the
individual subtractions.  The visit 12 subtraction was registered to the
position of the occulted star in the visit 11 subtraction and rotated to the
same celestial orientation. 

\subsection {Artifacts in PSF-Subtracted Images} \label{Artifacts}

Residual low spatial frequency systematic artifacts remain in the
combined image. These artifacts arise, primarily, from the interaction
of two effects: different centerings under the wedge of HR 4796A and the
reference PSF and a slight defocus (breathing excursion) in the Visit 12
PSF relative to HR 4796A image taken in the same visibility period.
They may be thought of as comprised of several ``features''.  Most
obvious among them are the radial ``spokes.''  We also see a radial
modulation of the background intensities, and a ``hemispheric'' bias in
that modulation due to decentering.  As a result, it is not possible to
simultaneously obtain zero nulling of the backgrounds at all radii with
the subtraction method described in \S~\ref{Subtractions}.  The
PSF-subtracted ``input'' images employed global nulling and minimization
in their creation, but they do not show ``optimal'' subtraction in all
radial zones.

The specific spatial nature of these residuals is complex.  Building a
multi-parameter (or multi-dimensional) model in the hopes of achieving
truly global background minimization is likely infeasible because many
instrumental effects, such as wavelength and position dependent wedge-
edge scattering that remain uncharacterized.  We accept that a global
fully-optimized PSF-subtracted image may not be achievable. 

\begin{figure}
\plotone{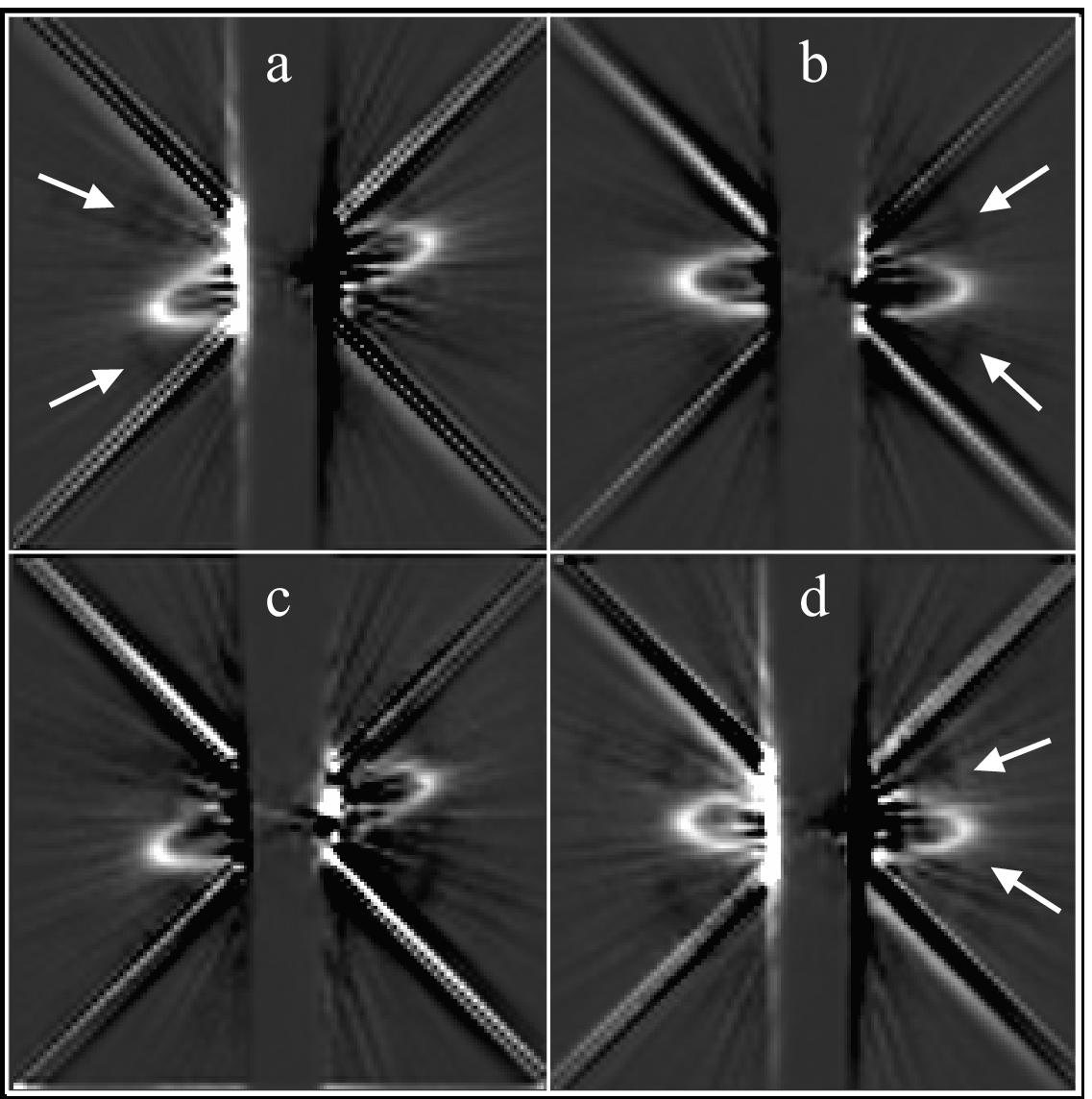}
\caption{PSF subtracted images of HR 4796A after PSF flux scaling and
  registering. The four panels show different target/PSF
  combinations. HR 4796A images from visit 11 are at the left ({\it a,c}) and
  visit 12 at right ({\it b,d}). The top row shows subtractions using the
  contemporaneous observations of HR 4748 (11$-$A1 for {\it a} and 12$-$A2 for
  {\it b}, and the bottom row shows the opposite pairs (11$-$A2 for {\it
    c} and 12$-$A1 for {\it d}).  The image scales and orientations are the
  same as in Figure 1. The arrows indicate some of the subtraction residuals
  discussed in section \ref{Artifacts}. On panel {\it a}, the arrows point to
  the negative zone near the NE ansa. On panel {\it b}, they point to the
  negative residuals near the SW ansa. On panel {\it d}, they show the
  positive residuals to the SW. \label{fig_intermedimages}}
\end{figure}

The NE side of the ring is less affected by residual diffraction and
instrumental scattering artifacts than the SW side of the ring.  The
appearance of the artifacts in the PSF-subtracted images may be better
understood by examining paired-subtractions of the individual HR 4748 PSFs
(Figure \ref{fig_intermedimages}).  Because the star positions in the two
visits were offset in X with respect to the wedge centerline, one edge of the
wedge appears bright, and the other dark after registration and subtraction.
The stellar positions were also offset in Y (along the wedge) changing, in
detail, the scattered light profile from the edge of the wedge.  Additionally,
a very small focus change occurred causing the asymmetrical appearance of the
+45\degr\ and -45\degr\ diffraction spike residuals.  Despite these
non-repeatabilities, the registration process mimics the minimization of the
radial streaks as was done for the HR 4796A PSF subtractions.

Figure \ref{fig_intermedimages} illustrates the nature and spatial
distribution of the subtraction residuals within 1$''$ of the star before
combination into the final image. Small differences between the individual PSF
subtracted images are seen.  Also shown are the azimuthally changing
influences of image artifacts within each PSF subtracted image.  For example,
due to the Visit A1 decentering of the HR 4748 reference PSF with respect to
the mid-line of the wedge, the resulting low-amplitude, large spatial scale,
negative residuals in the PSF subtraction are less pronounced on either side
of the NE ansa; i.e., locally, this is a ``cleaner'' PSF subtraction then
around the SW ansa.

The artifacts in individual subtractions are partially self-canceling in
weighted average combination, but unfortunately not completely so.  Note
in Figure \ref{fig_intermedimages} that the negative-going residuals at
radii just at and marginally beyond the SW ansa are of larger amplitude
in the subtractions using the Visit A2 reference PSF (panels b and c),
and more noticeable in panel b because of the lower amplitude of the
radial spokes as the HR 4796A-to-HR 4748 relative centering was very
precise.  A dark (negative) zone is seen abutting the NE ansa in the
subtractions of the Visit A1 PSF (panels a and d), where the reference
PSF star was significantly offset from the wedge center.  Finally, note,
that there is a zone of positive residuals seen in the panel d
subtraction to the SW just outside of the radius of the negative zone.

\section{Results} \label{Results}

\subsection{Ring Geometry \label{geometry}}

We exploit the high spatial resolution of the STIS data to improve upon
the initial determination of the geometrical parameters of the ring
derived from the NICMOS scattered-light images by \cite{Schneider1999}.
By ring geometry we mean: ({\it a}) the location of its center (determined
separately from the position of the star), ({\it b}) its ellipticity
(apparent eccentricity; we interpret the apparent ellipticity as evidence for a
disk which is intrinsically circular (or nearly so) and inclined to our
line of sight and use it to calculate an inclination), ({\it c}) the
major axis length - corresponding to the ridge of maximum surface
brightness in a deprojected ellipse, and ({\it d}) the major axis
celestial position angle of the apparent ellipse.

\begin{figure}
\plotone{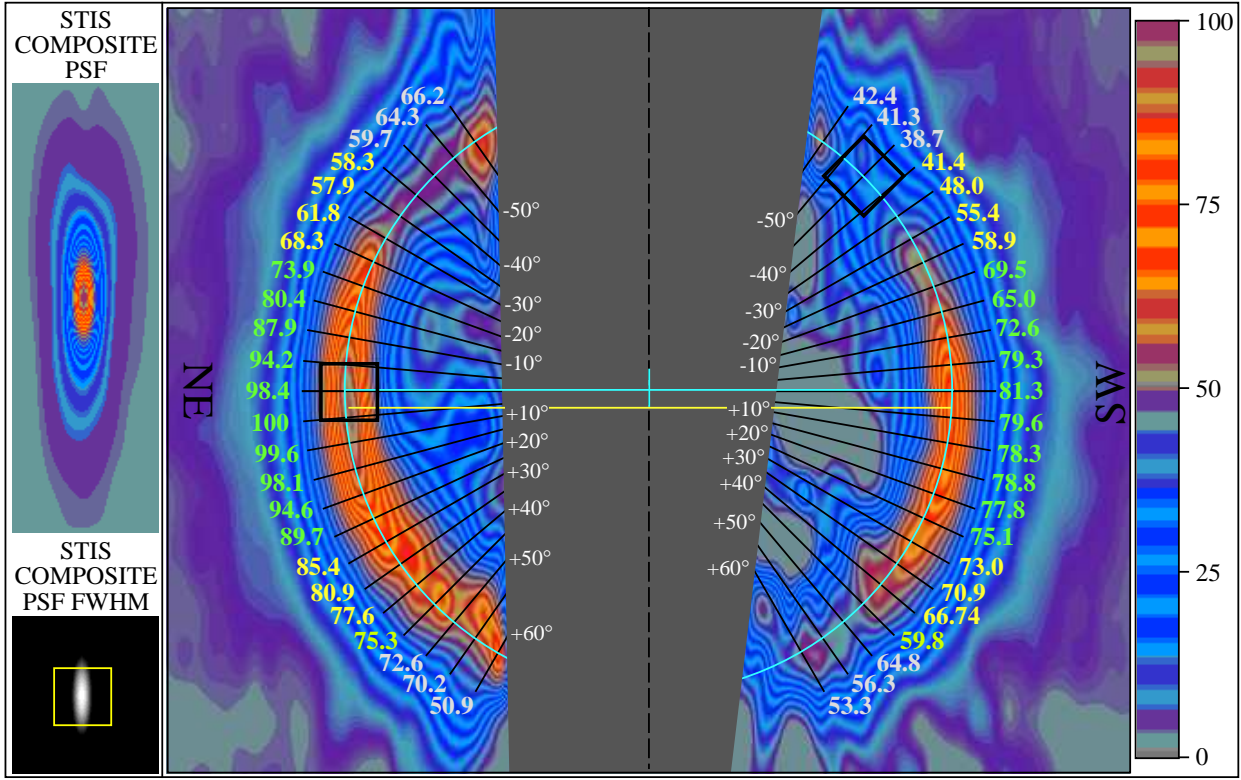}
\caption{Deprojected image of the disk, with flux conservation, showing flux
  isophotes. The color bar shows 3.2\% intensity isophotes switching from
  orange to blue at 50\% of peak intensity.  We fit concentric circles to
  seven ring isophotes at the 25, 50, 75, and 100\% flux density levels
  normalized at each azimuth to the maximum intensity at each position angle
  in each azimuthal bin.  Azimuthal samples were made at 5\degr\ deprojected
  increments as marked on the interior of the image. Photometry for these
  samples was done in apertures fully enclosing the projected FWHM of the STIS
  PSF, as shown in the bold black squares, which are rotated to maintain a
  fixed inner and outer distance from the ring center.  The numbers around the
  periphery of the disk image give the percentage of flux relative to the
  bright ansal peak at each fit point along the 100\% contour.
  The light blue circle shows the best fit at the 100\% contour.
 The horizontal blue line shows the ellipse major axis. The
  yellow line connects the ansae and therefore shows the offset of the ellipse
  due to forward scattering. Also shown at left are what the STIS PSF looks
  like when deprojected, (top) on a stretch that shows all of its details and
  (bottom) the same PSF clipped at the FWHM to show only the core shape. The
  yellow square on the lower PSF shows the size of the photometric extraction
  aperture.  \label{fig_deprojected}}
\end{figure}

\begin{figure}
\plotone{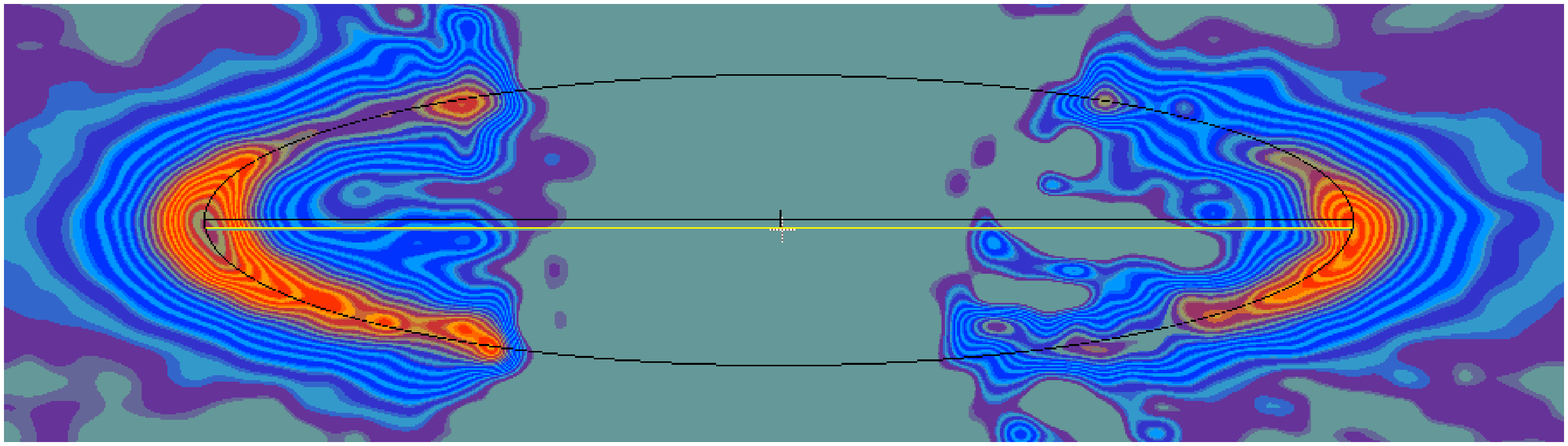}
\caption{Best fit ellipse (in black), determined from the contours as
  described in the text,  shown overplotted on the HR 4796
  image. The yellow line connects the ansae, and is parallel to the major
  axis of the ellipse to within the uncertainties. The vertical tick marks
  the center of the best-fit ellipse. \label{fig_fitellipse}}
\end{figure}

We determined the geometry of the ring in two ways:

First, we fit ellipses to the ring isophotes at the 25, 50, 75 (interior and
exterior to the peak), and 100\% flux density levels normalized at each
azimuth (in 5\degr\ deprojected increments; see Figure \ref{fig_deprojected})
to the maximum intensity at each position angle in each azimuthal bin.  In
doing so we exclude data behind and immediately adjacent to the coronagraphic
wedge and diffraction spikes. All of the geometrical parameters, except the
semi-major, axis were determined from the mean of the seven isophotal fits;
the semi-major axis length (which corresponds to a deprojected radius of
maximum surface brightness) was determined from the 100\% isophotal contour
alone (shown in Figure \ref{fig_fitellipse}). The seven fit ellipses were
concentric to within 6 mas.

Second, we built a model of the disk assuming an elliptical ring with a
surface brightness profile across the ring described by a Gaussian. The
model geometry is defined by 5 free parameters, corresponding to those
independently measured as described above: ellipse center (x and y),
semi-major axis length and orientation, and inclination.  The disk model
also includes two other free parameters; the ring Gaussian width and
overall intensity normalization (i.e., a scale factor to match the total
disk flux density).  For this purpose, we modeled the forward scattering
with a Henyey-Greenstein prescription (see \S 4.3). Finally, we modeled
the ansal asymmetry as a smooth sine function peaked along the major
axis at the NE ansa and decreasing to the SW side with an amplitude
equal to the photometrically measured asymmetry (see \S 4.4).  The model
free parameters were iteratively adjusted to fit the observational
data with chi-square minimization.

The geometrical parameters determined by the two independent methods
agree very well, and the averages from the two methods (and their
resulting uncertainties) are presented in Table \ref{tab_geometry}. In
Figure \ref{fig_fitellipse}, we show the best-fit ellipse (BFE, in black)
defined by the systemic parameters detailed in Table 2 overlaid on 3.2\%
photometric intensity contours (see scale bar in the figure).

\begin{deluxetable}{ll}
\tablecolumns{2}
\tablewidth{0pc}
\tablecaption{Ring Geometry \label{tab_geometry}}
\tablehead{}
\startdata
Position Angle of Major Axis           &27.01\degr   $\pm$ 0.16\degr \\
Major:Minor Axis Length Ratio          & 4.10  $\pm$ 0.05 \\
Implied Inclination                    & 75.88\degr   $\pm$ 0.16\degr \\
Ansal Separation (Photocentric Peaks)  & 41.563   $\pm$ 0.088 pixels  \\
                                       & 2\farcs107  $\pm$  0\farcs0045  \\
Major Axis (Best-Fit Ellipse [BFE])    & 41.698   $\pm$ 0.108 pixels \\
                                       & 2\farcs114  $\pm$  0\farcs0055 \\
Henyey-Greenstein $g$                  &0.16      $\pm$  0.06 \\
Ansal Asymmetry Factor                 &0.74      $\pm$  0.07 \\
\enddata
\end{deluxetable}

\subsection{Ring - Star Offset \label{offset}}

The stars themselves are hidden under the coronagraphic wedge during the
observations, so their positions cannot be measured directly.  Hence, we
determined the location of HR 4796A in each visit by using the
unocculted portions of the stellar diffraction spikes beyond the
coronagraphic wedge.  We assume the diffraction spikes are centered on
the star.  At every detector column, a slice was made through the
spikes, and the center of the spike was found by fitting a Gaussian
profile to the slice. A least squares fit was made to the coordinates of
the two orthogonal spikes, with the intersection defining the location
of the star.  The diffraction spikes are easily traced from a distance
of 9 pixels (0.46'') from the star (beyond the edge of the occulting
wedge) to 44 pixels (2.23'') at the edge of subarray used to read out
the image. The formal uncertainty on the fit location of the center is
0.08 pixels in each coordinate from propagating the uncertainty in the
two least squares fits.

The wedge-occulted positions of HR 4796A resulting from the
independent target acquisitions in Visits 11 and 12, as determined from
the above diffraction spike fitting method, were found to be (306.41,
41.28) $\pm$ (0.09, 0.06) pixels and (306.09, 40.01) $\pm$ (0.07, 0.08),
respectively, in the STIS CCD science instrument aperture frame (SIAF).
The inferred target positioning offset between the two visits is
therefore (+0.32,+1.27) $\pm$ (0.11, 0.10) pixels. This offset was
independently checked using the diffraction spike noise-minimization
iterative-subtraction method described in \S \ref{Subtractions} applied to the two HR
4796A visits; from which a position offset of (+0.26, +1.30) pixels was
found (i.e within 1$\sigma$ of the offset determined directly from
diffraction spike fitting).

We compare the position of the star to the geometrical center of the
ring ellipse to investigate the possibility of a stellocentric offset
such as seen in the case of the Fomalhaut debris ring as discussed by
\cite{Kalas2005}. 

\begin{figure}
\plotone{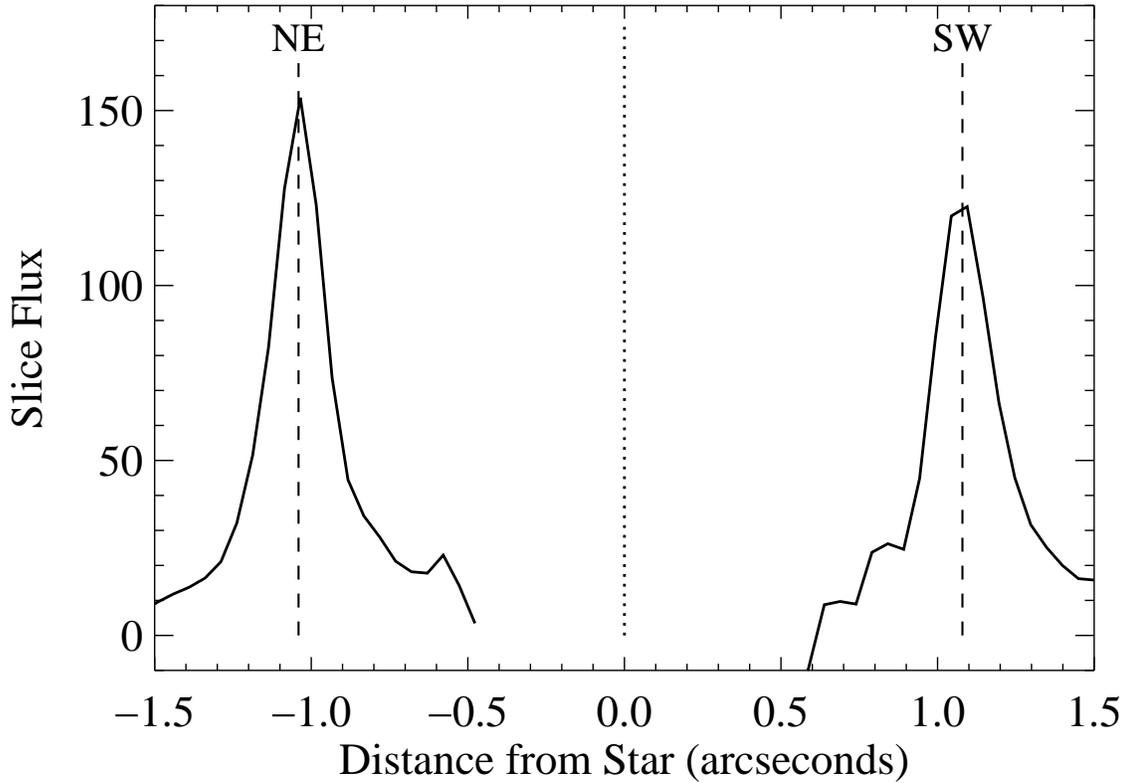}
\caption{A one pixel slice along the major axis of the disk shows the
  locations of the NE and SW ansae at distances of 1\farcs040 and 1\farcs079
  from the star respectively.  These locations are shown with the
  vertical dashed lines. The stellar location at zero is marked with a
  dotted line.  The point halfway between the ansae, i.e. the ring center,
  appears shifted 19 $\pm$ 6 mas away from NE ansa, where this uncertainty
  includes contributed both from fitting the locations of
  the ansae and determining the stellar center. At the distance to HR 4796 of 72.8 pc,
  this is a ring offset of 1.4 $\pm$ 0.4 AU.\label{fig_ringoffset}}
\end{figure}

We found the distance between the two ring ansae and the star by placing the
stellar center at the middle of a pixel and rotating the final image about the
center so the major axis was horizontal.  We took a one pixel slice along the
major axis and fit the locations of the ansae in this slice with Gaussians. We
find that the NE ansa is 20.522 $\pm$ 0.041 pixels (1\farcs040 $\pm$
0\farcs002) from the stellar center and the SW ansa is 21.287 $\pm$ 0.069
pixels (1\farcs079 $\pm$ 0\farcs003) from the stellar center, as shown in
Figure \ref{fig_ringoffset}. These uncertainties are the formal uncertainties
from the Gaussian fit and do not include the uncertainty in the center found
above. To make the ansae equidistant implies a ring center shifted away from
the NE ansa by 0.38 $\pm$ 0.12 pixels = 0\farcs019 $\pm$ 0\farcs006.  Because the
portion of the ring along its minor axis is obscured by the coronagraphic
wedge, we cannot robustly set a limit on a star-ring offset in this direction.

\subsection{Width of the Dust Ring \label{ring_width}}

\begin{figure}
\plotone{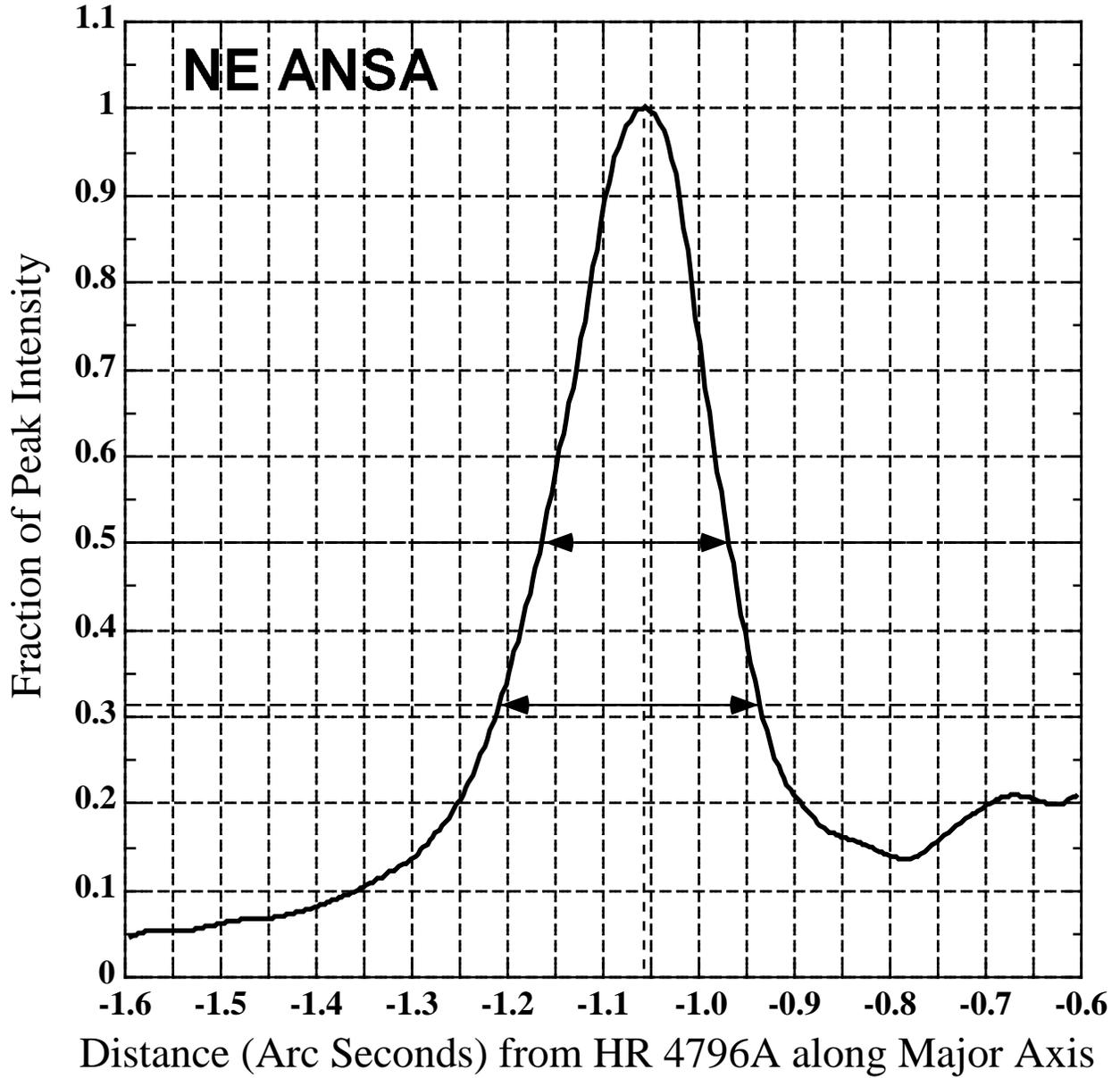}
\figcaption{Characteristic width of the HR 4796A ring defined from a
  cross-sectional profile through the NE ring ansa along the major axis
  of the best-fit ellipse.  The arrows show the
  cross-sectional profile's exterior and interior half widths
 at half intensity maximum and 1- 1/e of maximum. The distance along the
 major axis is measured from the center of the best-fit ellipse.\label{fig_profiles}}
\end{figure}

HR 4796A disk grains are radially confined to a narrow annular zone
circumscribing the star.  Radially across the NE (brighter) ansa we
measure a one pixel wide FWHM of the surface brightness profile as
0\farcs197, but the brightness fall-off is somewhat steeper interior to
the ring ansa then outward from it (see Fig \ref{fig_profiles}),
indicating a sharper interior cut-off than the outer truncation.  In
particular the measured HWHM interior to the NE ansa is 0\farcs089
whereas exterior to the ansa it is 0\farcs108. An azimuthally averaged
radial profile, extending in sectors $\pm$15$^\circ$ in elliptical
coordinates derived from the geometrical eccentricity of the ring, about
the ring ansae, is best fit to a broader Gaussian with a FWHM of
0\farcs222. Taking individual one-pixel wide profiles (in 5$^\circ$
deprojected increments) around the ring, the FWHM of the radial surface
brightness profiles is seen to be at a minimum through the ansae, and
increases in radial profiles closer to the ring minor axis. These
effects are opposite of what one would expect for a perfectly flat disk
and likely indicates scattering from dust above the mid-plane of the disk.

We characterize the width of the annular region confining most of the
starlight-scattering dust by the intrinsic FWHM of the ring at the
brighter ansa, which as measured is instrumentally broadened by the STIS
50CCD (unfiltered) point spread function.  Subtracting the FWHM of the
STIS 50CCD PSF (0\farcs070) in quadrature from the observationally
measured width, we find the intrinsic characteristic ring width to
be FWHM 0\farcs184 (13.4 $\pm$ 0.8AU, at the revised stellar parallax,
or 8.7\% of the ring diameter) with corresponding HWHM[inner] =
0\farcs082 and HWHM[outer] = 0\farcs102. This is consistent with the
upper limit set by Schneider et al. 1999 from the lower spatial
resolution NICMOS images.

\subsection{Forward Scattering \label{section_fs}}

For an inclined disk, the side centered on scattering phase angle
0$^\circ$ is seen primarily in forward scattered starlight while the
other half, centered on scattering phase angle 180$^\circ$, is seen
primarily in light that is backscattered.  The HR 4796A scattered-light
disk is clearly brighter on the east side (both north and south of the
minor axis) than on the west side.  To determine if the ring dust
displays significant azimuthally non-isotropic scattering, we measured,
and ratioed, the total flux density in symmetrically placed sectors
within the four readily visible quadrants of the disk -- between
5$^\circ$ and 20$^\circ$ from the major axis, and within 0\farcs8 to
1\farcs5 from the star.

We characterize this asymmetry in terms of a Henyey-Greenstein phase
function parameter, $g$, at the best fit inclination of 14.1$^{\circ}$
from edge-on.  The variable $g$ represents a measure of the
``front-to-back'' light-scattering asymmetry of the dust, where $g$=0
represents a completely isotropic phase function and $g$=1 represents
purely forward scattering.  The scattering angle, $\theta$ between an
incoming ray of light and the line-of-sight for an observer was
calculated following the formalism in Appendix A of \citet{mccabe2002}.
Making a Henyey-Greenstein model and integrating over the observed
scattering angles within each of the four photometric apertures
described above, using different $g$ values, we find a best fit for the
measured inter-sector flux density ratios with $g$=0.16 $\pm$ 0.06, or
evidence for forward scattering at the 2$\sigma$ level of confidence.

We also measured the ansa-to-ansa orientation angle, from the centroid of one
ansa to the other, as 26.6 $\pm$ 0.5 deg.  The difference between this orientation and the PA
of the best-fit disk ellipse (27.01 $\pm$ 0.16; Table \ref{tab_geometry}) is
0.4 $\pm$ 0.5 deg. That these position angles agree is another indication that
it is forward scattering that shifts the locations of the ansae away from the
disk semi-major axis (see yellow line in Figure \ref{fig_deprojected}).

\subsection{Ansal Asymmetry} \label{section_aa}

The NE side of the disk (flanking the brighter ansa) appears brighter
than the SW side. To measure this asymmetry, we took the total flux
density in the NE and SW ``lobes'', within $\pm$ 20$^\circ$ of the major
axis, and measured their ratios in those sectors between 0\farcs8 to
1\farcs5 from the star (using the same background annulus as for the
total disk photometry). The SW side is 0.74 $\pm$ 0.07 as bright as the
NE side.

\subsection{Photometry} \label{Flux}

Determining the flux density of the unobscured portion of the HR 4796A
ring is most affected by two sources of systematic uncertainties in the
measurements.  Intrinsically, the uncertainty in the brightness of the
ring in the STIS 50CCD instrumental system is dominated by optical
artifacts arising from the imperfect PSF subtractions (i.e., not by
photon statistics or other ``random'' noise sources).  Extrinsically,
effecting a photometric transfer from the instrumental system to a
standard photometric system (e.g., Johnson/Vega) is difficult because of
the extreme breadth of the 50CCD bandpass (FWHM = 75\% of the pivot
wavelength) and because the intrinsic SED / color
of the disk within that band is unknown.  We address these systematic
errors separately.

We measure the 50CCD bandpass disk flux density in an elliptical
aperture with the major:minor axis ratio and position angle as we
determined for the disk, and an inner semi-major axis of 0\farcs8 and an
outer semi-major axis of 1\farcs5. Inside 0$\farcs$8, pixel-to-pixel
fluctuations dominate the flux density measures on pixel size scales,
and outside 1$\farcs$5, light from the disk is less than 5-10\% of the
peak surface brightness.  Since only part of the disk was visible, we
restricted our aperture to be within 20$^\circ$ of either side of each
ansa.  We used an elliptical annulus between 2$\farcs$09 and 2$\farcs$50
with the same azimuthal coverage to determine the background level of
the image.

To estimate the systematic uncertainties introduced by PSF subtraction,
we create a grid of PSF-subtracted images in which we: (a) offset the
PSF template relative to the best determined position by 1-sigma of the
uncertainties in the X and Y (SIAF) co-aligned position determinations,
and (b) re-scale the intensity of the PSF template by 1-sigma from the
best determined target:template scaling ratio.  We use the resulting
standard deviation of the pixels in our photometric aperture, over all
the points in this grid, as an estimate of the per-pixel photometric
systematic uncertainty. We then apply this uncertainty to each pixel in
our photometric apertures.

To convert from raw instrumental units of counts~s$^{-1}$~pixel$^{-1}$
to per-pixel flux density in physical units of mJy under the STIS 50CCD
(unfiltered) passband, we adopt the instrumental sensitivity and
photometric zero point from version 7.0 (October 2003) of the STIS
Instrument Handbook for a spectrally flat source: AB magnitude 26.386 =
1 count~s$^{-1}$ (with an instrumental gain of 4.096 e$^-$~count$^{-1}$
as used for all our observations), with a zero-point magnitude of 3631
Jy.

The 50CCD band flux density of the visible portion of the disk
unocculted by the coronagraphic wedge and within the photometric
measurement aperture is then 5.47 $\pm$ 0.24 mJy (with the presumption
of a flat spectrum).  While this is likely to be very close to correct
for an A0V (B-V = 0.0) star, such as HR 4796A, the spectral reflectance
of the disk grains in the broad optical to near-IR is red
\citep{Debes2008} and will bias the flux density transformations into
standard photometric bands.  We use the {\it calcphot} task in the
STSDAS synphot synthetic photometry package (which provides a high
fidelity model of the STIS instrumental response) to estimate this
photometric uncertainty.  For disk grains as red as B-V = +0.60, the
V-band brightness of the disk (transformed from the broad 50CCD
passband) would be 0.047 magnitudes brighter then for grains with a
spectrally neutral (B-V = 0.0) reflectance; i.e., a 5\% error in the
absolute photometric calibration due to an uncertainty of this amount in
the B-V color of the circumstellar debris.

We compensate for the flux density originating from the unsampled
portion of the disk using the geometrical parameters in Table
\ref{tab_geometry}. We do this by building a model of the disk that
includes: (a) a Henyey-Greenstein phase function with the $g$ value
measured in Section \ref{section_fs}, (b) an ansal asymmetry
characterized by a sine dependence and a maximum excursion given in
Section \ref{section_aa}, (c) a width from the measured radial profile
given in Section \ref{geometry}, and (d) size from the measurements
described in Section \ref{geometry}. This model yields a total disk flux
density 1.71 times greater than in the measured region alone; if the
entire spatial region of the disk could have been measured, we would
therefore expect a 50CCD band total disk flux density of 9.4 $\pm$ 0.8
mJy.

\section{Discussion} \label{Discussion}

The $\sim$ 70 mas per resolution element STIS observation of the HR
4796A debris disk better elucidates the morphology and geometry of the
circumstellar ring than the longer wavelength NICMOS observations.
Most, if not all, of the ``clumpiness'' seen in the
\citet{Schneider1999} near-IR images of the ring was then suggested as
arising from optical artifacts in the NICMOS PSF subtractions, rather
than intrinsic to the dust distribution, and that indeed appears to be
the case.

\subsection{Ansal Asymmetry} \label{ASYM1}

A  hemispheric brightness asymmetry was suggested in both the NICMOS
F160W and F110W (1.6~$\mu$m and 1.1~$\mu$m) bands at the 10-15\% level
as noted by \citet{Schneider2001B}. A similar brightness asymmetry was
seen in the mid-IR and discussed by \citet{Telesco2000}. This work shows
the asymmetry in the broad-band optical to be about 25\%.

One model for the mid-infrared asymmetry was ``pericenter glow,'' a
forced eccentricity of the disk that resulted in one side's grains lying
closer to the star and therefore being heated to a higher temperature
\citep{Wyatt1999}.  To explain the mid-infrared observations of
\citet{Telesco2000}, required a radial offset asymmetry,
$(R_{SW}-R_{NE})/R_{mean}$ = 6.4\% \citep{Wyatt1999}.

We measure the offset between the center of the ring and the star along
the ring major axis to be
19 $\pm$ 6 mas or 1.4 AU (Figure \ref{fig_ringoffset}), for a radial offset
asymmetry of 3.7\%. Of course, the true displacement may include a
component toward or away from us. In that case, what we
interpret as forward scattering could be partially due to the east side
of the disk being closer to the star than the west side. The magnitude of
of such a shift is limited by the small asymmetry we measure.

In any case, the offset along the major axis of 3.7\% causes a brightness
asymmetry, just due to the $r^{-2}$ dependence of the stellar
brightness, of 14\%, which is approximately half of the observed
difference in the ansal brightnesses.

This asymmetry could also be due to an azimuthally inhomogeneous
population of disk grains, or arise from a non-uniform spatial
distribution of an intrinsically homogeneous population.  Given orbital
periods of $\sim$350 years for ring particles at the characteristic
radius of 76 AU, and an age of the star of $\sim$8 Myr, one would expect
the grains to be well mixed rather than azimuthally segregated
populations of grains with different scattering efficiencies.  We
suggest the possibility that the hemispheric variation in surface
brightness might arise, at least in part, from a dynamical redistribution of
the grains, through perturbations or resonant interactions with
co-orbital bodies, leading to azimuthally confined regions of enhanced
particle densities.  \citet{Wyatt1999} describe the effect of forced
eccentricities from the companion, which might be sufficient to explain a
portion of the asymmetry. \citet{KuchnerHolman03}, e.g. describe more generally how
planets can concentrate dust at resonances.

\subsection{An Intra-Ring ``Gap''?} \label{ASYM2}

\begin{figure}
\plotone{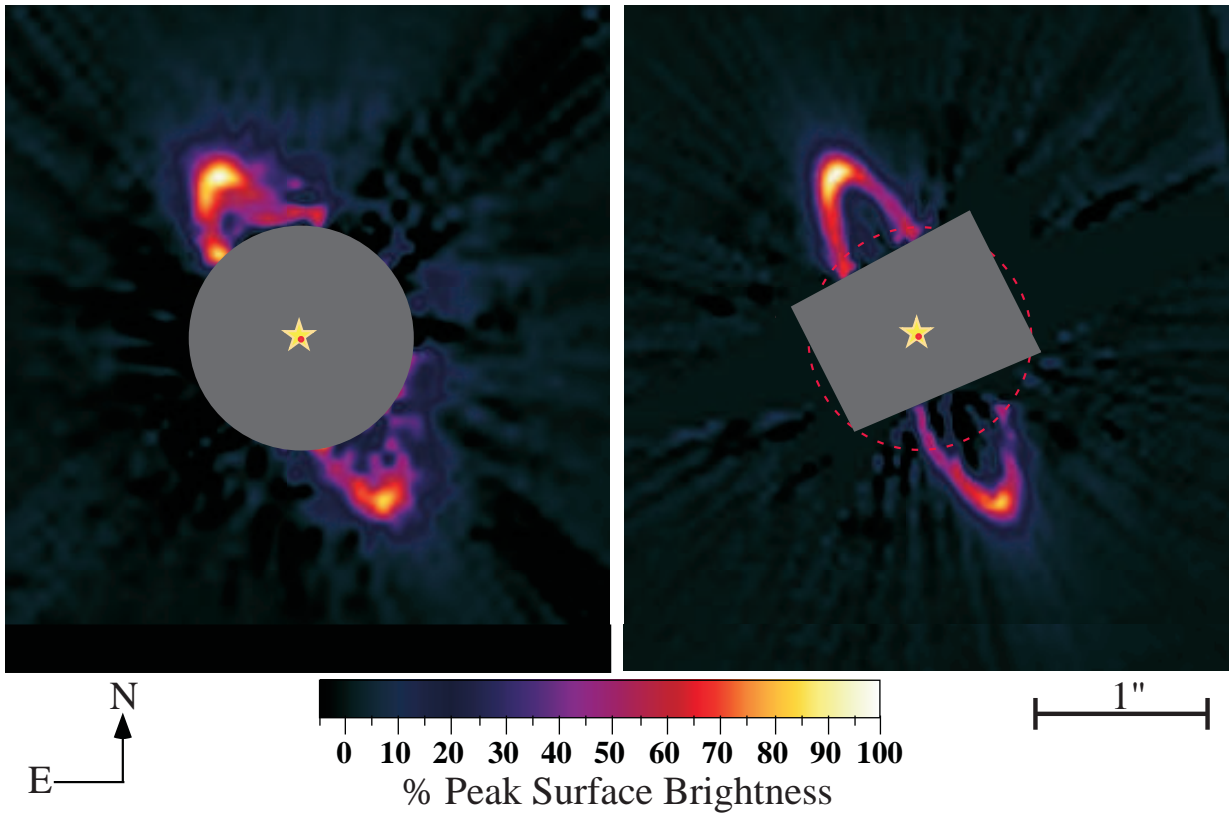}
\caption{Comparison of the NICMOS image of HR 4796A from Schneider et
  al. (1999) with the image from this work. For purposes of display and
  to highlight the morphology, the STIS image has been resampled
  2$\times$ (as was the NICMOS image) and low level background
  modulation from PSF subtraction at radii $<$ 1 arcsec has been
  estimated by Krieg interpolation and removed. The red dot marks the center
  of the ring while the yellow star marks the location of the star. Both images show a
  decrement in ring brightness at PA$\sim 230\degr$. \label{fig_nicmos_stis}}
\end{figure}

We note the appearance of a correlated decrement in the ring brightness at a
position angle (eastward from north) of 230$\pm$10\degr\ in both the NICMOS
F110W and STIS 50CCD images (Figure \ref{fig_nicmos_stis}).  The statistical
significance of this is not strong in the individual images; however, the
correlation of this ``feature'' in the two images, taken at significantly
different field orientations, wavelengths (PSF subtraction artifacts spatially
scale with wavelength), and instruments, is intriguing and suggests there
might be a gap in the distribution of ring particles at that azimuth.  This,
perhaps, is further evidence (albeit not very strong on its own) of azimuthal
sculpting of the grains by gravitational perturbation by one or more
undetected bodies of planetary mass.  This region is nearly opposite the
bright ansa, and is in the direction of the companion M-star that sits 7\farcs7
away at a PA = 225.4$^\circ$.

\subsection {``Confinement'' of the Disk Particles} \label{Confinement}

 The central ``hole'', seen to an inner radius of $\sim$ 0\farcs5 (35AU)
 in the 50CCD images is essentially cleared of circumstellar material
 which would otherwise scatter starlight.  A similar central clearing is
 apparent (as implied from the radial inward brightness fall-off) in the
 \cite{Schneider1999} 1.1 and 1.6 \micron\ NICMOS images.  Additionally,
 \citet {Telesco2000} find a mid-IR optical depth in the central zone of
 $<$ 3\% from 10.8 and 18.2~$\mu$m imaging. The disk's emitting regions
 are spatially coincident with the scattered light seen in the optical
 and near-IR.

Small particles may be cleared by radiation pressure and
Poynting-Robertson drag acting on timescales much shorter than the
estimated $\sim$ 8~Myr age of HR 4796A if the dust were not replenished
and/or confined.  The number of grains in the ring, as inferred from the
mid-infrared observations, implies that the dust in the ring should be
collisionally dominated. It is likely that mutual collisions
disintegrate large grains until they are small enough to be removed by
radiation pressure. Then, the absence of dust interior to the
starlight-scattering ring indicates a corresponding absence of large
bodies located there \citep{Kenyon1999}.

The outer ``truncation'' of the disk might be dynamically aided by HR 4796B, the
M-star companion to HR 4796A noted by \citet{Jura1995}, though it is at a
projected (minimum) distance of $\sim$ 500 AU.  Dynamical models of the
ring must also reproduce the relatively sharp inner, and slightly less
steep outer ring boundaries.  Such tight
bounding might be manifested by the ``shepherding'' of the ring
particles by co-orbital bodies \citep{Goldreich1979}, loosely analogous,
for example, to structures seen in the Saturnian rings, but here on
solar system size scales.

\subsection{Color (Spectral Reflectivity) of the Grains}
\label{Color}

We have found the HR 4796A disk grains to be significantly red at
optical to the near-IR wavelengths; this result is discussed in some
depth in a companion paper \citep{Debes2008}. Intrinsically red grains
are consistent with a collisionally evolved population of particles
where the characteristic particle size is larger than the wavelengths of
the observations, as opposed to neutral or blue grains which would be
more typical of a primordial population of ISM grains.

HR 4796A's disk is not the only one to display such red colors. The more
substantial disk around HD 100546 also shows red optical scattering
\citep{Ardila2007}, as do portions of the similarly thin $\beta$
Pictoris disk \citep{golimowski2006}. Perhaps the reddening mechanism is
common to both our outer Solar System and the cool portions of younger
disks around more luminous stars.

\acknowledgements

We gratefully acknowledge contributions from members of the STIS IDT
leading to a better understanding of some of the performance
characteristics of the instrument, and in particular Phil Plait for
assisting with some of the nuances of STIS data calibration, and Carol
Grady and Randy Kimball in regard to temporal stability and pointing
precision issues.  We also wish to thank other members of the NICMOS IDT
for their suggestions and comments during the investigatory phases of
this study, and specifically to Marcia Rieke for establishing the
absolute photometric calibration of NICMOS, and Elizabeth Stobie for
implementing significant improvements to the IDP3 software in support of
this study. This work is based on observations with the NASA/ESA Hubble
Space Telescope, obtained at the Space Telescope Science Institute,
which is operated by AURA, Inc. under NASA contract NAS2-6555 and
supported by NASA grants NAG5-3042 and GO 8624 and 10177.

\end{document}